# On Locality in Quantum General Relativity and Quantum Gravity

**Eduard Prugovečki**

*Department of Mathematics*
*University of Toronto*
*Toronto, Canada M5S 1A1*

**Abstract.** The physical concept of locality is first analyzed in the special relativistic quantum regime, and compared with that of microcausality and the local commutativity of quantum fields. Its extrapolation to quantum general relativity on quantum bundles over curved spacetime is then described. It is shown that the resulting formulation of quantum-geometric locality based on the concept of local quantum frame incorporating a fundamental length embodies the key geometric and topological aspects of this concept. Taken in conjunction with the strong equivalence principle and the path-integral formulation of quantum propagation, quantum-geometric locality leads in a natural manner to the formulation of quantum-geometric propagation in curved spacetime. Its extrapolation to geometric quantum gravity formulated over quantum spacetime is described and analyzed.

## 1. INTRODUCTION

The physical concept of locality in quantum physics has undergone a number of revisions during course of this century, as it became increasingly intertwined with that of causality. Its origins are, however, clearly rooted in the geometric conceptualization of "point," and of the topological concept of "neighborhood." In a recently advanced formulation of quantum general relativity[1] it has been shown that a formulation of this concept, which distinguishes between its various geometric and topological nuances, can provide the basis for a consistent unification of general relativity and quantum theory.

In the axiomatic approach to geometry, which culminated in Hilbert's work on the subject towards the end of the last century, the concept of "point," together with other key geometric concepts, is not defined. Thus, Hilbert himself starts his *Grundlagen der Geometrie* by requiring the reader to *imagine* "three kinds of things ... called points ... lines ... planes," and then enumerates the axioms to be fulfilled by these "things," and by the relationships between them.

On the other hand, Riemann's attitude to geometry, which influenced Einstein in his formulation of general relativity, is empirical. Thus, Riemann asserts that "the theorems of geometry cannot be deduced from general notions of quantity, but that those properties which distinguish Space from other conceivably triply extended quantities can only be deduced from experience." (Ref. 2, p. 135.) When such an attitude is adopted, geometry becomes of direct relevance to physics, and can be used as an analytic tool in physical theories.

In Sec. 2 we present an analysis meant to disentangle from each other the contemporary concepts of locality, microcausality and local commutativity of quantum fields. With the essential aspects of the concept of locality thus illucidated, we turn in Sec. 3 to the formulation of this concept in special relativistic quantum theory in Minkowski space. We explain why in that context the conventional formulation of this concept is deficient, and how it can be replaced by a consistent formulation incorporating a fundamental length. We then extrapolate in Sec. 4 this modified concept of locality to quantum theory in curved spacetime, and explain the distinction between its geometric aspects and its topological aspects. In Sec. 5 we turn to the formulation of the quantum bundles and of the local quantum frames required for the mathematical implementation of these ideas, and discuss the operational interpretation of the resulting fibre theoretical framework for quantum general relativity. Finally, in Sec. 6 we describe, in primarily nontechnical terms, how the ensuing concept of locality that incorporates a fundamental length (whose existence has been predicted by measurement-theoretical arguments) can be applied to the formulation of a geometric framework for quantum gravity which avoids the foundational difficulties encountered by various existing conventional frameworks for quantum gravity.

## 2. LOCALITY, MICROCAUSALITY AND LOCAL COMMUTATIVITY

As is well-known, in Euclidean geometry a "point" is defined as the intersection of two lines. With the concept of line *a priori* given, this definition throws no light on the physical nature of the objects that can be represented by points. Nevertheless, the notion of "material point" played a crucial role in Einstein's formulation of both special[3] and general[4] relativity.

Clearly, in adopting this notion Einstein relied on the "intuitive" grasp of this notion that one acquires with exposure to classical mechanics. There, however, this concept is viewed as an idealization. In this idealization, under certain circumstances the actual motion of macroscopic bodies is replaced with that of their center of gravity, with their spatial extension being thereby neglected.

When, however, the concept of material point plays a fundamental role in the formulation of the very concept of spacetime, then questions arise as to the physical legitimacy of this concept. Hence, in the collection[5] of essays honoring Einstein's seventieth birthday, H. Margenau voiced the criticism that "particles may not be regarded as points but as structures of finite size . . . , which threatens the validity of [relativistic] causal description" (Ref. 5, p. 259). This observation becomes extremely pertinent at the microlevel, when one tries to combine quantum theoretical notions with relativistic ones. In particular, it affects even QFT (i.e. quantum field theory) since, as emphasized by Bohr and Rosenfeld,[6,7] the question of localizability of quantum fields is inextricably intertwined with that of test bodies required for their measurement.

In conventional special relativistic QFT it has become customary to identify the notions of locality, microcausality and local commutativity of quantum fields. The proffered reasoning is that "local" measurements at two spacelike separated points $x$ and $y$ cannot affect each other, so that this presumed microcausality feature makes such measurements compatible; that, in turn, is supposed to imply that (integer-spin) quantum fields at those points have to commute. However, each step in this chain of reasoning can be brought into question.

First of all, as pointed out by Bogolubov *et al.*,[8] "we have no special reason for supposing that the measurement of a component of a Hermitian field at some point has no influence on the value of the components of this field at another point separated from the first point by an arbitrarily small spacelike interval." (Ref. 8, p. 373.) Second, as documented in a history[9] of this subject : "The *exact* meaning of this commutator relation is not clear. [T]his is underscored in the 1957 LSZ paper in which an operator $A(x)$ satisfying $[A(x), A(y)] = 0$ for spacelike separations [of $x$ and $y$] is *defined* as a *causal* operator – without discussing here (or anywhere in that paper) the physical interpretation. It seems *plausible* that it may have something to do with first-signal-principle [i.e., with Einstein causality], but the equivalence of these two definitions of causality is not established. The main justification of this new criterion for a causal theory is that it produces a desired result (dispersion relations), while not obviously conflicting with a more direct notion of causality." (Ref. 9, p. 214.)

However, dispersion relations are expressed in terms of scattering amplitudes and $S$-matrix elements. When one considers the manner in which the perturbative computations of these objects are performed in conventional QFT, one encounters yet another paradox: the local commutativity axiom plays no role in such computations.

Indeed, in the case of a Hermitian scalar free quantum field, we have that

$$[A(x), A(y)] = i\Delta(x - y) \quad , \qquad \Delta(x) = \Delta^{(+)}(x) + \Delta^{(-)}(x) \quad , \tag{1}$$

where the two-point "function" on the righ-hand side of (1) (which is actually a singular distribution) vanishes when $x$ and $y$ are spacelike separated. However, it is not *this* two-point "function" that enters in $S$-matrix computations, but rather the Feynman one

$$\Delta_F(x) = \theta(x^0)\Delta^{(+)}(x) - \theta(-x^0)\Delta^{(-)}(x) \quad . \tag{2}$$

As is well-known, this latter two-point "function" does not vanish when $x$ and $y$ are space-like separated, but rather decreases exponentially as that spacelike separation is increased.

Dirac was fully aware that the problem of QFT divergencies can be traced to the type of quantum field localizability which he introduced when he founded QED. Thus, he eventually arrived at the following conclusion: "Present-day atomic theories involve the assumption of localizability, which is sufficient but it is very likely to be too stringent. The assumption requires that the theory shall be built up in terms of dynamical variables that are each localized at some point in space-time, two variables localized at two points lying outside each other's light-cones being assumed to have zero P.b. [i.e., Poisson bracket]. A less drastic assumption may be adequate, e.g., that there is a fundamental length $\lambda$ such that the P.b. of two dynamical variables must vanish if they are localized at two points whose separation is space-like and greater than $\lambda$, but need not vanish if it is less than $\lambda$." (Ref. 10, p. 339.)

However, the axiomatic formulation of QFT subsequently introduced by Wightman[11] showed that local commutativity at all spacelike separations can be derived from the much weaker postulate that quantum fields commute only at very large spacelike separations, provided that other assumptions ordinarily made in conventional QFT are retained. Thus, such a (physically unwarranted) postulate as the existence of a global vacuum has to be also disposed off before a program based on Dirac's suggestion can be implemented.

An indication as to the general direction in which a formulation of a quantum theory incorporating a fundamental length is to proceed can be found in Born's observation that "the mathematical concept of a point in a continuum has no direct physical significance." (Ref. 12, p. 3). Born described this thesis in greater detail as follows:

"Statements like 'A quantity $x$ has a completely definite value' (expressed by a real number and represented by a point in the physical continuum) seem to me to have no physical meaning. Modern physics has achieved its greatest successes by applying a principle of methodology that concepts whose application requires distinctions that cannot in principle be observed are meaningless and must be eliminated. This is possible without difficulty in the present case also; we have only to replace statements like $x = \pi$ cm by: the probability of the distribution of values of $x$ has a sharp maximum at $x = \pi$ cm." (Ref. 13, p.167.)

Thus, the technical implementation of the idea that "the mathematical concept of a point in a continuum has no direct physical significance" can be achieved by regarding such points as representing only the *mean* locations of quantum events, and then supplementing that description of quantum events with additional information pertaining to the margins of uncertainty in their localization.

## 3. LOCALITY IN SPECIAL RELATIVISTIC QUANTUM MECHANICS

It is a measurement-theoretical fact that even in the classical regime no actual measurement of time or spatial location in relation to a Lorentz frame can be adequately described exclusively by four Minkowski coordinates, but has to be supplemented by the specifications of margins of confidence of the obtained readings. The dismissal of these margins of confidence in all theoretical considerations in *classical* special relativity is rooted in the belief that the accuracy of such measurements can be increased indefinitely, so that in a purely conceptual limit they approach zero. But is such an assumption justified at the quantum level?

The fact that actually it is not justified can be deduced from the conventional special relativistic framework for quantum mechanics.

For example, let us consider the case of a spin-0 particle of mass $m$. In the momentum representation space all its state vectors are elements of the Hilbert space $L^2(V_m^+)$ of wave functions on the forward mass hyperboloid $V_m^+$ that are square integrable with respect to the Lorentz invariant measure

$$d\Omega_m(k) = \delta(k^2 - m^2)\, d^4k \; . \tag{3}$$

This Hilbert space carries the inner product

$$\left\langle \tilde{\varphi}_1 | \tilde{\varphi}_2 \right\rangle = \int_{k^0 > 0} \tilde{\varphi}_1^*(k) \tilde{\varphi}_2(k)\, d\Omega_m(k) \; , \tag{4}$$

which is consistent with the interpretation of the values $\tilde{\varphi}(k)$ assumed by these wave functions as probability density amplitudes for measurements of 4-momenta, since $|\tilde{\varphi}(k)|^2$ gives rise to a probability density for any normalized $\tilde{\varphi}$.

No similar interpretation vis-à-vis space and time measurements can be given, however, to the values assumed by the configuration-space wave functions

$$\hat{\varphi}(x) = (2\pi)^{-3/2} \int_{k^o > 0} \exp(-i\,x \cdot k)\, \tilde{\varphi}(k)\, d\Omega_m(k) \ , \tag{5}$$

since the time evolution governed by the Klein-Gordon equation does not leave in this case the $L^2$-inner product invariant. Actually, as is well known, it is an inner product of the form given by

$$\langle \hat{\varphi}_1 | \hat{\varphi}_2 \rangle = i \int_\sigma \varphi_1^*(x)\, \overset{\leftrightarrow}{\partial}_\mu\, \varphi_2(x)\, d\sigma^\mu(x) \ , \tag{6}$$

that has to be adopted for the wave functions in (5) in order to secure the unitarity of the transformation from the momentum to the configuration representation.

The various attempts at finding physically acceptable position operators proved unsuccessful (cf. Ref. 14 for a survey). Amongst these attempts, the best-known was that of Newton and Wigner.[15] It was, however, declared[16] unsuccessful by Wigner himself after Hegerfeldt[17] gave a rigorous proof that any wave function (5) representing a state localized with respect to any family of self-adjoint position operators within a bounded spatial region will propagate instantaneously outside the causal future of that region, thus violating Einstein causality. This no-go theorem remains valid for arbitrary spins.

A similar negative outcome holds for probability currents. Thus, after statements appeared in literature to the effect that the Klein-Gordon current is a probability current when restricted to positive-energy solutions of the Klein-Gordon equation, Gerlach *et al.* proved[18] rigorously that the opposite is actually the case, namely that the timelike component of that current assumes for any such wave function negative as well as positive values at various points in Minkowski space. Furthermore, although in the case of spin-1/2 the Dirac current has a positive-definite timelike component, it turns out[19] that there are no positive-energy solutions of the Dirac equation that represent wave functions which are localized, at any time and with respect to any given global Lorentz frame, in any bounded region of space. In other words, no sharp localization can be achieved in that case either.

All of the above indicates that the notion of sharp localization, which works perfectly well in the nonrelativistic regime, becomes deficient in the relativistic regime, and should be therefore replaced by some form of unsharp localization. It turns out[20-23] that a formulation of unsharp localization emerges naturally when one considers representations of the (restricted) Poincaré group $\mathrm{ISO}_0(3,1)$ on the special relativistic phase $M^4 \times V^+$, where $M^4$ denotes the Minkowski space (in which we adopt the choice +1,–1,–1,–1 for the diagonal elements of the Minkowski metric $\boldsymbol{\eta}$), and $V^+$ is the forward 4-velocity hyperboloid.

A particularly simple realization of such a representation incorporating a fundamental length $\ell$ is provided by

$$U(a,\Lambda) : \ f(q - i\ell v) \ \mapsto \ f'(q - i\ell v) = f(\Lambda^{-1}(q-b) - i\ell\Lambda^{-1}v) \ , \tag{7a}$$

$$q \in M^4 \ , \qquad v \in V^+, \qquad (a,\Lambda) \in \mathrm{ISO}_0(3,1) \ , \tag{7b}$$

where, in Planck natural units, the wave functions

$$\varphi(\zeta) = \tilde{Z}_{\ell,m}^{-1/2} \int_{k^o > 0} \exp(-i\,\zeta \cdot k)\, \tilde{\varphi}(k)\, d\Omega_m(k) \ , \quad \zeta = q - i\ell v \ , \tag{8}$$

can be obtained, upon renormalization, by analytic continuations of those in (5). These wave functions constitute a Hilbert space with the inner product

$$\langle \varphi_1 | \varphi_2 \rangle = \int \varphi_1^*(\zeta)\varphi_2(\zeta)\, d\Sigma(\zeta) \ , \tag{9}$$

where the integration is carried out with respect to the unique (modulo a multiplicative constant) invariant phase space measure

$$d\Sigma(\zeta) = 2\, v^\mu\, d\sigma_\mu(q) d\Omega(v) \ , \qquad d\Omega(v) = \delta(v^2 - 1)\, d^4 v \ , \tag{10}$$

over any hypersurface $\sigma \times V^+$ in $M^4 \times V^+$ for which, as in case of (6), $\sigma$ is a spacelike Cauchy hypersurface in $M^4$. The value of the normalization constant in (8) can be expressed in terms of modified Bessel function $\mathrm{K}_2$ (cf. Ref. 1, Secs. 3.8 – 4.1 and Ref. 22, Sec. 2.9)

$$\tilde{Z}_{\ell,m} = 8\pi^4\, \mathrm{K}_2(2\ell m)/\ell m^2 \ , \tag{11}$$

and it is uniquely determined by the requirement that the inner product in (4) be equal to that in (9) under the map $\tilde{\varphi} \mapsto \varphi$ determined by (8). This secures the unitarity of that map, and therefore the unitary equivalence of the present phase space representation and the conventional momentum representation.

Due to the form (9) of the inner product, $\varphi(\zeta)$ in (8) can be consistently interpreted, in the context of measurement-theoretical schemes[24,25] dealing with simultaneous unsharp measurements of position and momentum, as the probability density amplitude for a measurement outcome, with an accuracy of the order $\ell$, of spacetime location $q$, and of 4-momentum $mv$ with a corresponding accuracy that complies with the uncertainty principle.

This interpretation is supported by the existence of the probability current

$$j^\mu(q) = 2 \int_{V_m^+} v^\mu \left| \varphi(q - i\ell v) \right|^2 d\Omega(v) \ , \tag{12}$$

which turns out to be conserved as well as relativistically covariant. Furthermore, in the nonrelativistic limit, obtained by using generic units and then letting $c \to +\infty$, the agreement of this interpretation with Born's interpretation in orthodox quantum mechanics can be established by taking the sharp-point limit $\ell \to 0$ (cf. Ref. 22, Sec. 2.6).

On the other hand, if the sharp-point limit $\ell \to 0$ is taken at a fixed and finite value of $c$, then divergences manifest themselves. Thus, at $c = 1$ in Planck units, we have

$$\tilde{Z}_{\ell,m} = (4\pi^4/\ell^3 m^4) + O(\ell^{-1}) \ \xrightarrow{\ \ell \to +0\ } \ +\infty \ . \tag{13}$$

The special relativistic regime does not provide any hints as to the value of the fundamental length $\ell$. However, measurement-theoretical studies[26,27], carried out for the case when gravity is present, clearly show that the Planck length $\ell = 1$ (equal to the Planck time in the adopted natural units) provides a lower bound on the value of $\ell$.

It might be thought that the formulation of a quantum spacetime framework incorporating a fundamental length $\ell$ would necessitate the introduction of a discrete structure for spacetime. This possibility has been actually considered already very early in the development of quantum theory, and has witnessed a recent revival of interest (cf. Ref. 1, Sec. 1.3 for references).

Historically, its best-known advocate was H. S. Snyder, who claimed[28] that the replacement of Minkowski space with a lattice structure obtained from the spectra of four "Hermitian" operators $x, y, z$ and $t$ does not violate Lorentz invariance, since "the spectra of the operators $x', y', z'$ and $t'$ obtained by taking linear combination of $x, y, z$ and $t$, which leave the quadratic form [giving rise to the Minkowski metric] invariant, shall be the same as the spectra of $x, y, z$ and $t$." (Ref. 28, p. 39.) In this context it was pointed out that the spectra of $x, y, z$ are the discrete subsets $\{n\ell : n = 0, \pm1, \pm2,...\}$ of the entire real line; whereas the spectrum of $t$ is continuous, and coincides with the entire real line.

On the other hand, in quantum theory it is not the operators representing observables that are themselves measurable, but rather their expectation values as well as those of their spectral measures, in general, and the eigenvalues giving rise to their spectra, in particular. However, the group which leaves invariant these joint spectra as well as the Minkowski quadratic form is not the Lorentz group since, for example, it obviously does not incorporate all spatial rotations. Hence, Lorentz covariance in the accepted sense of the word is actually violated, since, amongst all Lorentz transformations, only those would be physically acceptable which map the chosen set of lattice points in $\mathbf{R}^3$ into itself.

In the general relativistic regime, the principle of general covariance would be also violated by such a scheme, since the notion of diffeomorphism invariance would obviously become meaningless, and those coordinate systems would be favoured that are used when a Lorentzian spacetime spacetime continuum is "quantized" by associating operators with those coordinates.

Clearly, before submitting special and general relativity to such radical surgery, whereby its most basic principles are removed, one has to ask whether this kind of surgery is actually warranted. In this context, it has to be recalled that "quantization" does not necessarily entail "discreteness." Indeed, in nonrelativistic quantum mechanics position and momentum are quantized, and yet the position and momentum operators possess purely continuous spectra. Hence, the existence of a fundamental length can be taken into account not by replacing the spacetime continuum with a discrete spacetime structure, but rather by supplementing the continuum spacetime structure with additional features which reflect basic limitations on the precision of spacetime localization. The fibre bundle framework used in differential geometry provides the natural means for achieving such a goal.

## 4. LOCALITY IN GENERAL RELATIVITY

The first step in extending the results on unsharp localization from the Minkowski spacetime $M^4$ to any curved spacetime represented by a globally hyperbolic Lorentzian manifold $\mathbf{M}$ with metric $\boldsymbol{g}$ consists of replacing the global Lorentz frames in $M^4$ with moving local Lorentz frames in that Lorentzian manifold $(\mathbf{M}, \boldsymbol{g})$.

The concept of moving frame was originally introduced by H. Cartan with general relativity in mind. In modern fibre theoretical language, a moving frame is a section of the section $\boldsymbol{s}$ of the general linear frame bundle $GL\mathbf{M}$,[1,2] i.e., a smooth assignment of a linear frame $\boldsymbol{u}(x) \in GL\mathbf{M}$ to each point $x$ in an open set within the manifold $\mathbf{M}$.

In the presence of a Lorentzian metric $\boldsymbol{g}$, one can restrict one's attention to local Lorentz frames $\boldsymbol{u}(x) = \{\boldsymbol{e}_0(x), \boldsymbol{e}_1(x), \boldsymbol{e}_2(x), \boldsymbol{e}_3(x)\}$ belonging to the Lorentzian frame bundle $L\mathbf{M}(\boldsymbol{g})$, so that their elements constitute an orthogonormal basis in the tangent space $T_{\boldsymbol{x}}\mathbf{M}$ at each $x$, and so that $\boldsymbol{e}_0(x)$ is timelike. According to the original operational definition of Einstein,[3,4] such a Lorentz frame can be physically realized by assigning to $\{\boldsymbol{e}_1(x), \boldsymbol{e}_2(x), \boldsymbol{e}_3(x)\}$ three unit "rigid rods" at right angles to each other, and assigning to $\boldsymbol{e}_0(x)$ a "standard clock." However, as acknowledged by Einstein in later years, the concept of "rigid rod" is inconsistent with the basic relativistic tenet which stipulates the impossibility of transmitting signals with supraluminal velocities.

Therefore, we shall adopt an alternative operational definition in which the spatial frame labeled by the triple $\{\boldsymbol{e}_1(x), \boldsymbol{e}_2(x), \boldsymbol{e}_3(x)\}$ is physically provided by a pointlike and neutral classical test particle that marks its origin $O$, and by six other identical test particles $A_a$, $a = \pm 1, \pm 2, \pm 3$, which are in its immediate vicinity, and mark its positive and negative axes. The verification as to whether such physical setup, taken in combination with "standard clocks" in the immediate vicinity of all these test particles, constitutes a local Lorentz frame consists of checks, under identical *local* conditions, of whether all the test particles $A_a$, $a = \pm 1, \pm 2, \pm 3$, are indeed at a unit distance from $O$, and whether the axes joining $O$ to each $A_a$, $a = \pm 1, \pm 2, \pm 3$, are at right angles to each other. In principle, such operational procedures can be carried out by observing the recoil of light signals sent between the test particles constituting such a classical frame. In CGR (i.e., classical general relativity) their constituents are assumed to behave deterministically, so that absolutely precise compensations for the momenta transferred by such recoils are deemed to be in principle feasible.

In special relativistic quantum theory it is not just the Lorentz group $SO_0(3,1)$, but the Poincaré group $\mathrm{ISO}_0(3,1)$ incorporating also spacetime translations, that plays a crucial role as a spacetime symmetry group. In the general relativistic regime these groups assume the role of gauge groups. Hence, the Lorentz group acts from the right[1,2] upon the local Lorentz frames $\boldsymbol{u}(x) \in L\mathbf{M}(\boldsymbol{g})$, so that, as one makes the transition from one moving Lorentz frame to another, families of Lorentz transformations $\Lambda(x)$ give rise to gauge transformations.

Similarly, the Poincaré group acts from the right upon the local Poincaré frames $\boldsymbol{s}(x)$,

$$(b(x), \Lambda(x)) \; : \; \boldsymbol{s}(x) \; \mapsto \; \boldsymbol{s}'(x) = \boldsymbol{s}(x) \cdot (b(x), \Lambda(x))^{-1} \; , \tag{14a}$$

$$\boldsymbol{s}(x) = (\boldsymbol{a}(x), \boldsymbol{e}_i(x)) \; , \quad \boldsymbol{s}'(x) = (\boldsymbol{a}'(x), \boldsymbol{e}_i'(x)) \; , \qquad x \in \mathbf{M} \; , \tag{14b}$$

as one makes the transition from $\boldsymbol{s}$ to another Poincaré moving frame $\boldsymbol{s}'$. These Poincaré moving frames represent cross-sections of the Poincaré frame bundle $P\mathbf{M}(\boldsymbol{g})$ over a Lorentzian spacetime manifold $(\mathbf{M}, \boldsymbol{g})$. In turn, this principal bundle consists of all the Poincaré frames $\boldsymbol{u} = \{(\boldsymbol{a}, \boldsymbol{e}_i) | \, i = 0, 1, 2, 3\}$ above all base locations $x \in \mathbf{M}$, obtained by translating in the tangent space $T_x\mathbf{M}$ in the amount represented by the 4-vector $\boldsymbol{a}$ a future-oriented and spatially right-handed local Lorentz frame $\{\boldsymbol{e}_i | \, i = 0, 1, 2, 3\}$ with origin at the point of contact of the tangent space $T_x\mathbf{M}$ and $\mathbf{M}$.

We note that in all these considerations the frames in question are deemed to be local in the sense that, at the mathematical level, their elements belong to tangent spaces $T_x\mathbf{M}$, rather than to $\mathbf{M}$ itself. Hence, in general, in the presence of a fibre bundle over the base manifold $\mathbf{M}$, we shall refer to all objects belonging to the fibres of that bundle as being *geometrically* local. This form of locality has to be distinguished from what we shall call *topological* locality, which pertains to objects in a neighborhood of a given point in $\mathbf{M}$.

Both these forms of locality occur in general relativity, but they are often confused in literature. For example, it has been pointed out that various "formulations of the principle of equivalence characteristically obscure [the] crucial distinction between first-order laws and second-order laws by blurring the distinction between 'infinitesimal' laws, holding at a single point, and local laws, holding on a neighborhood of a point." (Ref. 29, p. 202.) In the above introduced terminology, "infinitesimal" laws, holding at a single point, pertain to geometric locality; whereas laws holding on a neighborhood of a point pertain to topological locality.

This distinction enables the formulation of quantum mechanical theories with a fundamental length into frameworks that are local in the geometric sense, but not in a topological sense which entails that certain probability amplitudes vanish outside small neighborhoods of points in the base manifold $\mathbf{M}$.

To arrive at a better understanding of the role and significance of general relativistic geometric locality, let us consider first the classical case where the base Lorentzian manifold $(\mathbf{M}, \boldsymbol{g})$ is the Minkowski space $M^4$. From the point of view of general relativity, a 4-vector $\boldsymbol{X}$ representing such a physical entity as a 4-velocity belongs not to $\mathbf{M}$, but to some tangent space $T_x\mathbf{M}$. Hence, such a 4-vector represents a geometrically local entity, and has no global significance. Let us, however, choose in $M^4$ a global Lorentz frame $\mathcal{L} = \{\boldsymbol{e}_i(x_0) \mid i = 0,1,2,3\}$ with its origin at a point $x_0 \in \mathbf{M} \cong M^4$, as well as a global Poincaré moving frame given by the following cross-section of $P\mathbf{M}(\boldsymbol{g})$,

$$\boldsymbol{s}_0(\mathcal{L}) = \left\{ (\boldsymbol{a}(x),\boldsymbol{e}_i(x)) \middle| \ \boldsymbol{a}(x) = -x^i\boldsymbol{e}_i(x) \in T_x\mathbf{M}, \ \ x = x^i\boldsymbol{e}_i(O) \in \mathbf{R}^4 \right\}, \tag{15}$$

where $-\boldsymbol{a}(x)$ is the position 4-vector of each $x \in \mathbf{M} \cong M^4$ in relation to the origin at $x_0$. In this Poincaré moving frame all the fibres $T_x\mathbf{M}$ of the tangent bundle $T\mathbf{M}$ can be identified with its typical fibre $M^4$ in a manner which in CGR underlies the transition from the general relativistic to the special relativistic regime: each local vector $\boldsymbol{X} \in T_x\mathbf{M}$ is identified with a 4-vector in $M^4$ which has in relation to $\mathcal{L}$ the same components as $\boldsymbol{X}$ has in relation to $(\boldsymbol{a}(x),\boldsymbol{e}_i(x))$. Hence, a transition is achieved from the general relativistic and local point of view to the special relativistic and global point of view.

Conceptually the same type of construction and transition can be implemented in the quantum regime by introducing the concepts of quantum bundle and of local quantum frame.

## 5. QUANTUM BUNDLES AND LOCAL QUANTUM FRAMES

In general, a fibre bundle $\mathbf{E}$ that is associated to a principal bundle can be represented by a $\mathbf{G}$-product of the principal bundle and the typical fibre $\mathbf{F}$ of that associated bundle, where $\mathbf{G}$ is the structure group of the bundle – namely, in physics terminology, its gauge group. Thus, if the principal bundle is the Poincaré frame bundle $P\mathbf{M}(\boldsymbol{g})$ over a base Lorentzian manifold $(\mathbf{M},\boldsymbol{g})$, then by the definition of such a $\mathbf{G}$-product (cf. Ref. 1, Sec. 4.1),

$$\mathbf{E} = P\mathbf{M}(\boldsymbol{g}) \times_{\mathbf{G}} \mathbf{F} \ , \qquad \mathbf{G} = \mathrm{ISO}_0(3,1) \ , \tag{16}$$

its generic element $\boldsymbol{\Psi}$ above a base location $x \in \mathbf{M}$ is represented by the equivalence class

$$\boldsymbol{\Psi} = \{(\boldsymbol{u},\Psi) \cdot g^{-1} \mid g = (a,\Lambda) \in \mathrm{ISO}_0(3,1)\} \ , \qquad (\boldsymbol{u},\Psi) \in P\mathbf{M}(\boldsymbol{g}) \times \mathbf{F} \ , \tag{17}$$

within the Cartesian product $P\mathbf{M}(\boldsymbol{g}) \times \mathbf{F}$ of $P\mathbf{M}(\boldsymbol{g})$ and $\mathbf{F}$. For example, the tangent bundle $T\mathbf{M}$ over $(\mathbf{M},\boldsymbol{g})$ can be viewed as being the $\mathbf{G}$-product of $P\mathbf{M}(\boldsymbol{g})$ and $M^4$, so that in that case each equivalence class $\boldsymbol{\Psi}$ in (17) represents a 4-vector $\boldsymbol{X} \in T\mathbf{M}$.

In general, the equivalence class in (17) results from the action from the right $g : \boldsymbol{u} \mapsto \boldsymbol{u} \cdot g$ of the elements $g$ of the structure group $\mathbf{G}$ upon the Poincaré frames $\boldsymbol{u}$ within the fibre of $P\mathbf{M}(\boldsymbol{g})$ above $x$, and from the action from the left of a representation $U(g)$ of $\mathbf{G}$ upon the elements $\Psi$ of the typical fibre $\mathbf{F}$. For example, if $\mathbf{F}$ is the Hilbert space with the inner product in (9), and $U(a,\Lambda)$ is the representation in (7), then we shall have in (17) that

$$(\boldsymbol{u}, \Psi) \cdot g^{-1} = (\boldsymbol{u} \cdot (a, \Lambda)^{-1}, U(a, \Lambda) \Psi) \ . \tag{18}$$

In that case we shall refer to the associated bundle in (16) as a quantum bundle over $(\mathbf{M}, \boldsymbol{g})$.

The above construction gives rise to the unitary maps

$$\boldsymbol{\sigma}_x^{\boldsymbol{u}} : \ \Psi \ \mapsto \ \Psi \in \mathbf{F} \ , \qquad \Psi \in \mathbf{F}_x \ , \tag{19}$$

between the quantum fibre $\mathbf{F}_x$ above a given base location $x \in \mathbf{M}$ and the typical fibre $\mathbf{F}$. For each choice of Poincaré frame $\boldsymbol{u}$ above the same $x \in \mathbf{M}$ we shall refer to $\Psi$ as the *coordinate wave function* of the local state vector $\boldsymbol{\Psi} \in \mathbf{F}_x$, and to the map in (19) as a *soldering map*.

By the manner of their construction, the quantum state vectors $\boldsymbol{\Psi}$ provided by (17) are geometrically local. To understand their physical significance, we shall introduce in the quantum bundle $\mathbf{E}$ local quantum frames soldered to a Poincaré frame $\boldsymbol{u} \in P\mathbf{M}$.

If $\boldsymbol{u}$ lies above $x \in \mathbf{M}$, then such a frame is mathematically provided by the family

$$\mathcal{Q}_\ell^{\boldsymbol{u}} = \left\{ \boldsymbol{\Phi}_\zeta^{\boldsymbol{u}} \in \mathbf{F}_x \ \middle| \ \zeta = q - i\ell v \in \mathbf{C}^4, \ q \in \mathbf{R}^4, \ v \in V^+ \right\} \tag{20}$$

of generalized coherent states,[1,23] whose coordinate wave functions are represented by

$$\Phi_\zeta(\zeta') = \left( \boldsymbol{\sigma}_x^{\boldsymbol{u}} \boldsymbol{\Phi}_\zeta^{\boldsymbol{u}} \right)(\zeta') = K^{(\ell, m)}(\zeta'; \zeta) \tag{21}$$

in terms of the reproducing kernel

$$K^{(\ell, m)}(q'' - i\ell v''; q' - i\ell v') = \tilde{Z}_{\ell, m}^{-1} \int_{V_m^+} \exp\left\{ [i(q' - q'') - \ell(v' + v'')] \cdot k \right\} d\Omega_m(k) \ , \tag{22}$$

corresponding to the Hilbert space $\mathbf{F}$. This reproducing kernel has the property that

$$\Psi(\zeta) = \int K^{(\ell, m)}(\zeta; \zeta') \ \Psi(\zeta') d\Sigma(\zeta') \tag{23}$$

for all $\Psi \in \mathbf{F}$, and is also such that it displays the two basic properties of propagators,

$$K^{(\ell, m)}(\zeta''; \zeta') = K^{(\ell, m)*}(\zeta'; \zeta'') = \int K^{(\ell, m)}(\zeta''; \zeta) K^{(\ell, m)}(\zeta; \zeta') d\Sigma(\zeta) \ , \tag{24}$$

with the integrations being carried out as in (9). Due to these properties, the quantum frame in (20) provides a continuous resolution of the identity $\mathbf{1}_x$ in the fibre $\mathbf{F}_x$ to which its elements belong:

$$\int \left| \boldsymbol{\Phi}_\zeta^{\boldsymbol{u}} \right\rangle d\Sigma(\zeta) \left\langle \boldsymbol{\Phi}_\zeta^{\boldsymbol{u}} \right| = \mathbf{1}_x \ . \tag{25}$$

Hence, any local state vector $\boldsymbol{\Psi} \in \mathbf{F}_x$ can be expanded as follows:

$$\boldsymbol{\Psi} = \int \Psi(\zeta) \boldsymbol{\Phi}_\zeta^{\boldsymbol{u}} d\Sigma(\zeta) \ , \qquad \Psi(\zeta) = \left( \boldsymbol{\sigma}_x^{\boldsymbol{u}} \boldsymbol{\Psi} \right)(\zeta) = \left\langle \boldsymbol{\Phi}_\zeta^{\boldsymbol{u}} \middle| \boldsymbol{\Psi} \right\rangle \ . \tag{26}$$

Due to the fact that[1,23]

$$\boldsymbol{\Phi}_\zeta^{\boldsymbol{u}} = U(q, \Lambda_v) \boldsymbol{\Phi}_{-i\ell\hat{v}}^{\boldsymbol{u}} \,, \qquad \zeta = q - i\ell v \,, \qquad \hat{v} = (1, \mathbf{0}) \,, \tag{27}$$

the quantum frame associated by (20) with the classical Lorentz frame $\boldsymbol{u}$ can be operationally obtained by considering a quantum test body marking the origin of that frame, and subjecting it to the family of transformations that describes the basic kinematical operations of spacetime translation and boost that occur in (27). Verifications as to which of these test bodies mark orthonormal axes can be carried out in the manner described in the previous section for classical Lorentz frames, under the proviso that the compensatory mechanisms adopted in Sec. 4 are now subject to the uncertainty relations. This implies that the orientations of the axes of quantum frames are operationally determined only stochastically, and that all their mutually orthogonality relations hold only in the mean.

In the case where the base Lorentzian manifold $(\mathbf{M}, \boldsymbol{g})$ is the Minkowski space $M^4$, the transition from general relativistic local quantum states to special relativistic global quantum states can be carried out as described in Sec. 4 for the classical case. Thus, in the quantum general relativistic context, a state vector $\boldsymbol{\Psi} \in \mathbf{F}_x$ represents a geometrically local entity, and has no global significance. If, however, we choose in $M^4$ a global Lorentz frame $\mathcal{L}$, then in the Poincaré moving frame (15) all the fibres $\mathbf{F}_x$ of the quantum bundle $\mathbf{E}$ can be identified with its typical fibre $\mathbf{F}$. This is achieved by identifying each local state vector $\boldsymbol{\Psi} \in \mathbf{F}_x$ with the state vector $\Psi \in \mathbf{F}$ represented, in relation to the quantum frame corresponding to $\mathcal{L}$, by the coordinate wave function of $\boldsymbol{\Psi} \in \mathbf{F}_x$ in relation to the local quantum frame (20) corresponding to $(\boldsymbol{a}(x), \boldsymbol{e}_i(x))$ in (15). In this manner the transition is made from the general relativistic quantum-geometric framework introduced in this section to the special relativistic framework described in Sec. 3.

On the other hand, no such transition is possible when the base Lorentzian manifold $(\mathbf{M}, \boldsymbol{g})$ is curved. In that case, as in CGR, in the present formulation of QGR (i.e., quantum general relativity) only the local point of view is acceptable. In particular, in view of the interpretation of the left-hand side of (8) as a probability amplitude, and of the manner in which quantum frames are operationally defined, one can assign to the coordinate functions in (21) the interpretation of local quantum metric fluctuation amplitudes that affect the Riemann normal coordinates generated, in a suitably small neighborhood of each $x \in \mathbf{M}$, by the exponential map determined by the orthonormal frame $\{\boldsymbol{e}_i(x) | \, i = 0, 1, 2, 3\} \in L\mathbf{M}(\boldsymbol{g})$.

Indeed, due to the strong equivalence principle, an operational meaning can be assigned in CGR to such coordinates: they are inertial normal coordinates in those sufficiently small neighborhoods in which second-order derivatives of the metric tensor can be neglected. Hence, in any such neighborhood, and to that order, the probability densities

$$\left| \left\langle \boldsymbol{\Phi}_\zeta^{\boldsymbol{u}(x)} \middle| \boldsymbol{\Psi} \right\rangle \right|^2 \,, \qquad \zeta = q - i\ell v \,, \tag{28}$$

supply for any local quantum state vector $\boldsymbol{\Psi} \in \mathbf{F}_x$ the relative probabilities of measurement outcome values $q^i$ for the normal coordinates of points in that neighborhood, as well as of values $v^i$ for the stochastic 4-velocity components with respect to the local quantum frame soldered to a Poincaré frame $\{(\mathbf{0}, \boldsymbol{e}_i(x)) | \ i = 0,1,2,3\} \in P\mathbf{M}(\boldsymbol{g})$ which is part of an inertial moving frame. In particular, if we choose $\boldsymbol{\Psi}$ equal to a state vector in a local quantum frame (20), then we arrive at the conclusion that (21) supplies locally the relative probability density amplitudes for the stochastic fluctuations in the measurement verifications of the spacetime location of the origin of that quantum frame. In fact, by an adaptation[30] to the quantum-geometric framework of Wigner's measurement-theoretical treatment[31] of the time-energy uncertainty relation, it follows that these fluctuations manifest themselves equally well in timelike as in spacelike directions. This is in accordance with the most basic relativistic tenets. In this manner, geometrically local features of the present framework for QGR are translated into topologically local ones.

In the present formulation of QGR the operation of state preparation has to be envisaged locally, and as resulting in some local state vector $\boldsymbol{\Psi} \in \mathbf{F}_x$. A physical mechanism is then required that governs the propagation of that local state vector to various locations in $\mathbf{M}$, and is such that, in the special relativistic regime where $(\mathbf{M}, \boldsymbol{g})$ is the Minkowski space $M^4$, it leads to the same outcome as the propagation governed by the Klein-Gordon equation.

This mechanism is provided by the quantum-geometric propagation described in detail in Ref. 1 as well as in Ref. 32. This type of propagation represents an extrapolation of Feynman's path integration method of formulating quantum propagation in nonrelativistic or in relativistic flat spacetimes.

According to the original formulation of this method,[33,34] quantum propagation takes place over limits of broken polygonal paths. These paths sequentially connect points in a foliation of such a spacetime into spacelike hyperplanes, and all the observed probability transition amplitudes result from the superposition of spacetime propagators over such paths.

Hence, the quantum-geometric extrapolation of this idea to a curved spacetime requires the foliation of the globally hyperbolic spacetime manifold $(\mathbf{M}, \boldsymbol{g})$ representing that spacetime into a family of Cauchy reference surfaces $\Sigma_t$. It then involves replacing straight lines with arcs of geodesics of the Lorentzian metric $\boldsymbol{g}$, and deriving spacetime propagators by extending the strong equivalence principle from the classical to the quantum regime. This means that free-fall quantum-geometric propagation is governed by the Levi-Civita connection determined by $\boldsymbol{g}$, so that the most fundamental features of classical general relativity are retained in the mean in the quantum regime. On the other hand, the fact that infinitesimally this propagation is governed by the special relativistic propagator in (24) which, in turn, converges in the sharp-point limit $\ell \to 0$, upon appropriate renormalization, to the Feynman propagator in (2), ensures that, when the curvature effects are negligible, the quantum-geometric propagation agrees with the conventional special relativistic propagation.

The physical meaning of quantum-geometric propagation emerges from a straight-forward extrapolation of the interpretation of (8) as a probability density amplitude along reference hypersurfaces in Minkowski space, as well as from an extrapolation of the interpretation of (28) as a local relative probability density amplitude. Thus, let the quantum-geometric propagation of the initially prepared local state vector $\boldsymbol{\Psi}_{x(t_0)}$ result in families of local state vectors $\boldsymbol{\Psi}_{x(t)}$ along subsequent reference hypersurfaces $\Sigma_t$, $t > t_0$, in a curved spacetime manifold $(\mathbf{M}, \boldsymbol{g})$. For any choice of Poincaré moving frame, corresponding to some cross-section $\boldsymbol{s}$ of $P\mathbf{M}(\boldsymbol{g})$, we can interpret

$$\Psi(x(t), v) = \left\langle \boldsymbol{\Phi}^{\boldsymbol{s}(x(t))}_{\hat{\zeta}(x(t))} \,\middle|\, \boldsymbol{\Psi}_{x(t)} \right\rangle \ , \qquad \hat{\zeta}(x(t)) = -a(x(t)) - i\ell v \ , \qquad (29)$$

as a quantum-geometric wave function, whose values along each subsequent reference hyper-surface $\Sigma_t$ provides the .i.relative probability amplitudes for the detection of the mean stochastic 4-velocity value $v$ at the base location $x(t) \in \Sigma_t$. Thus, if a local boson state represented by $\boldsymbol{\Psi}_{x(t_0)}$ were prepared at $x(t_0) \in \Sigma_{t_0}$, then the squares of the absolute values of the quantum-geometric wave function in (29) provide the relative probability densities that the boson will be found, by measurements of stochastic position and momentum performed along the reference hypersurface $\Sigma_t$ with test bodies of the same zero spin and rest mass $m$, to be located at $x(t)$ and have a mean 4-velocity $v$ in relation to the local quantum frame at $x(t)$. Naturally, these considerations can be generalized to arbitrary spins and rest masses of the system as well as of the test bodies.

It has to be emphasized that, generically, the amplitudes in (29) give rise only to *relative* probability densities. This, together with the fact that in the QFT context the present approach requires only local rather than global vacuum states, enables the avoidance of the foundational difficulties encountered by the conventional formulation of QFT in curved spacetime, which entails *ex nihilo* pair creation that is not part of the energy-conserving and well established phenomenon of pair creation (cf. Ref. 35, Sec. 3.3).

Indeed, in ordinary quantum mechanics in a single Hilbert space, the probabilities for spatial detection are absolute rather than conditional, since the wave functions in configuration space are defined globally at all times. However, when a local point of view is adopted, the preparatory procedures have to be envisaged locally, so that the resulting probabilities for future detection are conditional. In a classical stochastic process, such as Brownian motion, the conditional probabilities can be consistently normalized after the instant of preparation since, in principle, observations can be carried out which do not disturb the system. This is no longer the case in the quantum regime, since the process of measurement of position induces the "collapse" of the wave function. Hence, due to the combined effects of locality and "collapse" of the wave function, there is no justification for assuming that the probabilities of subsequent detection have to be normalized over all later reference hypersurfaces. Indeed, each *actual* position measurement leads to a reduction of the quantum-geometric wave function. This results in a local state, and therefore leads to a new set of conditional probabilities.

## 6. SUPERLOCALITY IN GEOMETRIC QUANTUM GRAVITY

The two principal types of approaches to quantum gravity, namely the covariant and the canonical, do not dwell on questions dealing with the meaning and role of locality in the presence of a quantum gravitational field. The reason for that can be traced to the methodologies of these two types of approaches.

The covariant approach to quantum gravity, which used to be popular in the 1970s, assumes an *a priori* given classical spacetime background $(\mathbf{M}, \boldsymbol{g}^{\text{back}})$. The quantum gravitational field $\boldsymbol{g}^{\text{quan}} = \boldsymbol{g}^{\text{back}} + \boldsymbol{g}^{\text{cor}}$ is then assumed to be globally defined on $\mathbf{M}$, with the quantum gravitational correction "correction" $\boldsymbol{g}^{\text{cor}}$ to the classical background field $\boldsymbol{g}^{\text{back}}$ being provided by gravitational radiation in the form of multi-graviton states defined globally on $(\mathbf{M}, \boldsymbol{g}^{\text{back}})$. The most common choice for $(\mathbf{M}, \boldsymbol{g}^{\text{back}})$ is the Minkowski space, whose metric properties obviously do not represent macroscopically the mean values of a realistic model of curved spacetime, and therefore provides no clues to the question of locality in quantum gravity. By concentrating on perturbative *S*-matrix methods based on renormalization methods which ultimately proved unsuccessful,[36] the covariant method leaves the question of locality unanswered beyond the formal manipulations characteristic of such methods. Furthermore, by simply adopting a fixed classical spacetime background, this method surrenders diffeomorphism invariance. Indeed, according to the active interpretation of general covariance,[37] if $\psi : \mathbf{M} \to \mathbf{M}$ is any diffeomorphism of a differential manifold $\mathbf{M}$ onto itself, then the two Lorentzian manifolds $(\mathbf{M}, \boldsymbol{g})$ and $(\mathbf{M}, \boldsymbol{g}')$ provide *physically* equivalent models of spacetime in case that $\boldsymbol{g}' = \psi^* \boldsymbol{g}$, i.e., if the two metrics are diffeomorphically equivalent. Therefore, the "covariant" method actually violates key aspects of the general covariance principle, which is crucial to the notion of locality in CGR.

The canonical approach to quantum gravity, which has witnessed a revival of interest during the past ten years, has emerged from the ADM canonical formulation of classical gravity. This latter formulation views spacetime as a geometrodynamic evolution of 3-geometries, provided by a 3-dimensional manifold $\Sigma$ and by Riemannian 3-metrics $\boldsymbol{\gamma}$ on $\Sigma$ which depend on a parameter $t$. For certain choices of initial conditions this parameter can play the role of global time. On the other hand, the formal viewing of the components of $\boldsymbol{\gamma}$ and their canonical conjugates (or, more recently, of certain "new variables"[37] giving rise to them) as providing an uncountable family of dynamical variables to be canonically quantized creates fundamental difficulties with the notions of "observable" and of "time." However, instead of considering the notions of space and time, which are essential to the notion of locality, as truly fundamental, and therefore as absolutely essential ingredients of a quantum gravitational framework from the outset, the various schools of canonical quantization consider the formal schemes they advance as being the actually most fundamental input, out of which everything else is supposed to emerge. Faced with the loss of the clear-cut notion of what is "time" in such a framework, the attitude of some of the recent proponents of these schemes is to simply state: "forget time" (Ref. 36, pp. 135 and 169). Naturally, under such circumstances, the question of locality in the quantum gravitational regime becomes irrelevant.

Thus, the proponents of covariant as well as canonical approaches share a belief in the primary significance of *formal* schemes of quantization, albeit there is no general consensus on the fundamental features which these schemes should possess.

Epistemologically as well as methodologically, this attitude is diametrically opposed to that of Einstein in founding general relativity, whereby the final outcome[4] resulted from a ten-year search for the right mathematical framework into which his fundamental physical ideas on the nature of space and time could be fitted, rather from a series of attempts to endow an *a priori* chosen formal framework with physical meaning.

The recently advanced geometric framework for quantum gravity[1] strives to adapt Einstein's methodology to the quantum regime by revising his epistemology in accordance with fundamental quantum mechanical tenets. Of course, this means that, contrary to Einstein's own beliefs, the determinism of CGR has to make way for the indeterminism characteristic of quantum theory. Hence, in geometric quantum gravity, all spatio-temporal relationships are stochastic in nature. The metric fluctuations entailed in the notion of localizations in relation to the quantum frames described in the preceding section supplies a partial source for that stochasticity. In addition, since changes in the quantum states of the matter fields affect the geometry of spacetime, that geometry can no longer be deterministically derived from the initial conditions.

Hence, although the geometric approach to quantum gravity assumes that the quantum gravitational field $\mathbf{g}^Q = \mathbf{g}^M + \mathbf{g}$ is represented by mean values provided by 4-metrics $\mathbf{g}^M$ and quantum gravitational fluctuations around those values due to the presence of gravitons to which the quantum gravitational radiation field $\mathbf{g}$ gives rise, those mean metrics are not *a priori* given, as is the case with the background metrics in the covariant approach; rather, they are generated, together with the underlying manifold structure, by the quantum geometrodynamic evolution of the quantum gravitational field in perpetual interaction with the "matter" fields – by which we mean, as customary, all quantum fields describing matter and nongravitational radiation. In this context, the diffeomorphism invariance present in CGR is extrapolated to QGR by regarding these fields as Grassmann-valued superfields over a supermanifold representing quantum spacetime.

To see qualitatively how this comes about, let us first consider the canonical algorithmic schemes[38] for generating a classical spacetime from an initial 3-metric $\boldsymbol{\gamma}_0$, extrinsic curvature $\boldsymbol{K}_0$ and "matter" fields $\boldsymbol{\phi}_0$ provided along a 3-manifold $\Sigma_0$. These data, obeying appropriate constraint equations, provide the starting point of the algorithm for the iterative computation, from geometrodynamic versions of the Einstein equations, of the values $\boldsymbol{\gamma}_n$, $\boldsymbol{K}_n$ and $\boldsymbol{\phi}_n$ of these same quantities within subsequent "thin" segments $\mathbf{S}_n$ $n = 0, \pm 1, \pm 2,...$, corresponding to parameter values $t \in [t_{n-1}, t_n]$. The classical spacetime manifold $(\mathbf{M}, \boldsymbol{g})$ is obtained from this geometrodynamic evolution in the limit $\varepsilon = t_n - t_{n-1} \to 0$. Its Lorentzian metric is related to the computed 3-metrics $\boldsymbol{\gamma}$ as follows,

$$\boldsymbol{g} = dt \otimes dt - \gamma_{ab}\, dx^a \otimes dx^b \ , \quad a, b = 1, 2, 3 \ , \tag{30}$$

if the initial data are provided in the form

$$g_{00} = 1 \ , \qquad g_{a0} = g_{0a} = 0 \ , \qquad g_{ab} = -\gamma_{ab} \ , \qquad \mathrm{K}_{ab} = -\tfrac{1}{2}\dot{g}_{ab} \ . \tag{31}$$

In that case the resulting parameter plays the role of proper time for a coherent flow of test particles whose worldlines are orthogonal to the resulting reference hypersurfaces. These synchronous hypersurfaces then provide a foliation of $(\mathbf{M}, \boldsymbol{g})$.

As described in detail in Chapter 8 of Ref. 1, in the quantum regime the aforementioned initial conditions provided along $\Sigma_0$ by the 3-metric and the extrinsic curvature in (31) have to be supplemented by a specification of the states of quantum gravitational radiation and quantum "matter" fields in superfibres above a supermanifold extension of $\Sigma_0$. Then, in gross outline, the quantum geometrodynamic evolution proceeds as follows: the presence of a specific state of the quantum gravitational field $\mathbf{g}^{\mathrm{Q}}$ within each base-segment $\mathbf{S}_n$ governs the quantum-geometric evolution of all the quantum "matter" fields from the inflow hypersurface $\Sigma_{t_{n-1}}$ to the outflow hypersurface $\Sigma_{t_n}$ of that base-segment; $\Sigma_{t_n}$ then becomes the inflow hypersurface of the next base-segment $\mathbf{S}_{n+1}$; the values of the local quantum energy-momentum density operators in the resulting states of the "matter" fields along the hypersurface $\Sigma_{t_n}$ determine the formation of the state of the quantum gravitational field $\mathbf{g}^{\mathrm{Q}}$ within the base-segment $\mathbf{S}_{n+1}$, by metrizing of the base-segment of $\mathbf{S}_{n+1}$ (in the sense of singling out a gauge orbit of diffeomorphically equivalent Lorentzian metrics over $\mathbf{S}_{n+1}$), and by singling out within the quantum gravitational fibres above its points $x$ the multi-graviton states of quantum gravitational radiation. This iterative process then repeats itself from segment to segment, with the limit $\varepsilon = t_n - t_{n-1} \to 0$ being eventually taken as in the earlier described classical case.

The presence of quantum gravitational constraints is dealt with by extending the base-segments $\mathbf{S}_n$ into supermanifolds $\mathbf{S}_n$. These supermanifolds emerge from the action of a quantum gravitational gauge supergroup whose elements correspond to Poincaré gauge transformations as well as to diffeomorphisms giving rise to physically equivalent representations of a quantum spacetime related to various physically equivalent mean values $\overline{\mathbf{g}^Q} = \mathbf{g}^M$ of the quantum gravitational superfield $\mathbf{g}^Q$. Thus, locality is amplified into a superlocality which requires the specification of Poincaré frame in addition to the location of the fibre in which it lies. This feature can be traced to the geometric interpretation of BRST symmetries in Yang-Mills theories as vertical automorphisms over principal bundles,[39] and it reflects the fact that the little (i.e., stability) group for massive representations of the Poincaré group merges, upon contraction, into that for mass zero representations in a manner which is not invariant under Lorentz boosts (cf. Ref. 1, Sec. 7.1).

All this shows that the notion of locality in geometric quantum gravity displays features that are not present in either CGR or in quantum-geometric theory over curved spacetime, since the latter adopts an *a priori* given base Lorentzian manifold, and therefore does not make allowances for the influence of the *quantum* "matter" fields upon the geometry of spacetime. It is the mutual interaction between *quantum* geometry and *quantum* "matter" that brings to the fore the superlocality features displayed by geometric quantum gravity.

## 7. CONCLUSION

The developments in conventional QFT in Minkowski space are usually described as being founded on a physical idea of "locality" to which the only alternative is the notion of "nonlocality" of the type first advocated in Yukawa's formulation of QFT, and subsequently further developed by Efimov and others.[40] These two types of formulations of QFT distinguish themselves primarily by the manner in which the interactions between quantum fields are formulated: in "local" QFT they are envisaged as taking place at single points $x$ in Minkowski space; whereas in "nonlocal" QFT a "smearing" is carried out, so that quantum fields over entire neighborhoods of each point $x$ in Minkowski space are deemed to partake in the mutual interaction.

On the other hand, it has been rigorously proved that, under the assumptions ordinarily made in conventional QFT, quantum fields at a given point $x$ in Minkowski space actually do not exist (cf. Ref. 41, Sec. 10.4), but rather that a "smearing" with test functions is required. However, the "constructive" QFT program[42] meant to produce models based on such a formulation of quantum fields ended up with the physically unacceptable conclusion that key 4-dimensional QFT models, such as QED, are trivial in the sense that their $S$-matrix are equal to the identity (cf. Ref. 42, p.120). On the other hand, the conventional treatment of such models, yielding highly nontrivial predictions well-confirmed by experiments, concentrated exclusively on formal perturbative $S$-matrix techniques in which, as pointed out in Sec. 2, locality does not actually play any physically or mathematically meaningful role even if interpreted as microcausality or local commutativity.

However, we saw in Sec. 2 that there is no well-founded justification for identifying the three concepts of locality, microcausality and local commutativity. Furthermore, our analysis in Sec. 4 revealed that the concept of locality displays already in CGR various geometric and topological nuances, which become totally submerged if such an identification is attempted in QGR. On the other hand, once these distinctions are pinpointed, we showed in Sec. 5 that some basic aspects of locality in CGR can be transferred to a quantum-geometric framework for QGR, where a fundamental length is present.

This demonstrates that the introduction of a fundamental length in quantum theory does not necessarily entail "nonlocality." In fact, if the concept of locality in relativistic quantum theory were interpreted correctly, rather than merely being borrowed from the classical regime without any deeper analysis of its meaning, then the opposite is true. Indeed, the presence of a fundamental length mediates[1] the unambiguous formulation in the QFT over curved spacetime of geometrically local physical quantities, such as stress-energy tensors, or energy, momentum and angular momentum densities, which are as essential to QGR as they are to CGR. This is exactly the opposite to the situation in the conventionally "local" formulation of QFT in curved spacetime. There, one encounters fundamental difficulties with the formulation of the stress-energy tensor for quantum fields even in the absence of mutual quantum field interactions, and one has to resort for its definition to renormalization schemes which proved to be only partly successful (cf. Ref. 35, Sec. 6.6).

All this indicates that the concept of locality is richer in its meaning and its implications than ordinarily surmised. This paper is meant to demonstrate that, if this concept is treated with proper regard for its range of possible mathematical and physical interpretations, it can provide the resolution to long standing problems in relativistic quantum theory.

It is a pleasure to contribute this paper to an issue of *Foundations of Physics* honoring the sixty-fifth birthday of Professor L. P. Horwitz, for he, in his well-known work[43,44] based on the Stueckelberg equation, has demonstrated an exceptional willingness to look beyond the immediate, the obvious and the fashionable.

## ACKNOWLEDGEMENT

This work was supported in part by the NSERC Research Grant No. A5206.